\documentclass[aps,pre,twocolumn,groupedaddress,amsmath,amssymb]{revtex4-1}
\usepackage{graphicx,color}
\begin{document}

%\preprint{}
\title{From polymers to proteins: effect of side chains and broken symmetry in the 
formation of secondary structures within a Wang-Landau approach}

\author{Tatjana \v{S}krbi\'{c}}
%\thanks{}
\affiliation{Dipartimento di Scienze Molecolari e Nanosistemi,
Universit\`{a} Ca' Foscari di Venezia,
Campus Scientifico, Edificio Alfa,
via Torino 155,30170 Venezia Mestre, Italy. E-mail: tatjana.skrbic@unive.it}
%%%%%%%%%%%%%%%%%%%%%%%%%%%%%%%%%%%%%%%%%%%%%%%%%%%%%%%%%%%%%%%%%%%%%%%%%%%%%%%%%%%
\author{Artem Badasyan}
\affiliation{Material Research Laboratory, University of Nova Gorica, 
SI-5270 Ajdovscina, Slovenia. E-mail: artem.badasyan@ung.si}
%%%%%%%%%%%%%%%%%%%%%%%%%%%%%%%%%%%%%%%%%%%%%%%%%%%%%%%%%%%%%%%%%%%%%%%%%%%%%%%%%%%
\author{Trinh Xuan Hoang}
\affiliation{Center for Computational Physics, Institute of Physics, Vietnam Academy 
of Science and Technology, 10 Dao Tan St., Hanoi, Vietnam. E-mail: hoang@iop.vast.ac.vn}
%%%%%%%%%%%%%%%%%%%%%%%%%%%%%%%%%%%%%%%%%%%%%%%%%%%%%%%%%%%%%%%%%%%%%%%%%%%%%%%%%%%
\author{Rudolf Podgornik}
\affiliation{Department of Theoretical Physics, J. Stefan Institute and Department of Physics,
Faculty of Mathematics and Physics,
University of Ljubljana - SI-1000 Ljubljana, Slovenia. E-mail: rudolf.podgornik@ijs.si}
%%%%%%%%%%%%%%%%%%%%%%%%%%%%%%%%%%%%%%%%%%%%%%%%%%%%%%%%%%%%%%%%%%%%%%%%%%%%%%%%%%%%
\author{Achille Giacometti}
\affiliation{Dipartimento di Scienze Molecolari e Nanosistemi,
Universit\`{a} Ca' Foscari di Venezia,
Campus Scientifico, Edificio Alfa,
via Torino 155,30170 Venezia Mestre, Italy. E-mail: achille@unive.it}

\date{\today}

\begin{abstract}
We use micro-canonical Wang-Landau technique to study the equilibrium properties of a 
single flexible homopolymer where consecutive monomers are represented by impenetrable 
hard spherical beads tangential to each other, and non-consecutive monomers interact 
via a square-well potential. To mimic the characteristic of a protein-like system, 
the model is then refined in two different directions. Firstly, by allowing partial 
overlap between consecutive beads, we break the spherical symmetry and thus provide 
a severe constraint on the possible conformations of the chain. Alternatively, we 
introduce additional spherical beads at specific positions in the direction normal 
to the backbone, to represent the steric hindrance of the side chains in real proteins.
Finally, we consider also a combination of these two ingredients. In all three systems, 
we obtain the full phase diagram in the temperature-interaction range plane and find 
the presence of helicoidal structures at low temperatures in the intermediate range of 
the interactions. The effect of the range of the square-well attraction is highlighted, 
and shown to play a role similar to that found in simple liquids and polymers. 
Perspectives in terms of protein folding are finally discussed.
\end{abstract}

\pacs{}
%\keywords{}

\maketitle

\section{Introduction}
\label{sec:introduction}
Square-well (SW) potential has a long and venerable tradition in simple liquids 
\cite{Hansen86}, and has become a paradigmatic test-bench for more sophisticated  
new approaches. In early attempts of numerical simulations of liquids \cite{Adler59}, 
it was initially used as a minimal model, alternative to Lennard-Jones potential, 
because it could be more easily implemented  in a simulation code, and yet contained 
the salient features of a pair potential for a liquid. It displays both a gas-liquid 
and liquid-solid transitions in the phase diagram, with results often quantitatively 
in agreement with real atomistic fluids \cite{Barker76}. For sufficiently short-range 
attraction, the gas-liquid transition becomes metastable and gets pre-empted by a direct 
gas-solid transition \cite{Hagen94,Pagan05,Liu05}. Several variants of the SW model have 
also been proposed over the years in the framework of molecular fluids \cite{Gray84} 
and colloidal suspensions \cite{Lyklema91}.

In the framework of polymer theory, the model is relatively less known, but it has 
experienced a renewed interest in the last two decades because it exhibits a reasonable 
compromise between realism and simplicity \cite{Zhou97,Taylor95,Taylor03,Taylor09_a,Taylor09_b}. 
In this model, the polymer is formed by a monomer sequence of impenetrable hard-spheres, so that 
consecutive monomers are tangent to one another, and non-consecutive monomers additionally interact 
via a square-well interaction. The model can then be seen in a perspective as a variation of the 
usual freely-jointed-chain \cite{Grosberg94}, with the additional inclusion of a short-range 
attraction and excluded volume interactions between different parts of the chain.

In spite of its simplicity, this model displays a surprisingly rich phase behavior, 
including a coil-globule and a globule-crystal transitions, that are the strict 
analog of the gas-liquid and liquid-solid transitions, respectively. Interestingly, 
a direct coil-dense globule transition is found for sufficiently short range attraction, 
pushing the analogy with the direct gas-solid freezing transition even further \cite{Taylor95,Taylor09_a}.

The SW polymer model, henceforth referred to as model P (Polymer), can in principle be 
refined to mimic the folding of a protein, rather than the collapse of a polymer. However, 
there are some crucial differences between synthetic polymers and proteins that should be 
taken into account. The first difference stems from the specificity of each amino acid 
monomer forming the polypeptide chain, that provides a selectivity in the intra-bonding 
arrangement. As amino acids are often classified according to their polarities,
one simple way of accounting for this effect at the minimal level is given by the 
so-called HP model \cite{Yue92}, where the selectivity is enforced by partitioning the 
monomers in two classes, having {\sl hydrophobic} and {\sl polar} characters.
Under the action of a bad solvent and/or low temperatures, the hydrophobic monomers will 
tend to get buried inside the core of the globule, in order to prevent contact with the 
aqueous solvent. The HP model has been shown to be very effective in on-lattice studies 
\cite{Wust08,Seaton09,Wust11}, to describe the folding process at least at a qualitative level. 
An alternative possible route that has been recently explored is the use of patchy interactions 
\cite{Coluzza11}.

Another crucial difference, that will feature as a focus of the present study, stems from the 
observation that a chain, composed only of spherical beads backbone, is unable to capture the 
inherent anisotropy induced by the presence of side chains \cite{Maritan00,Banavar03,Banavar09}. 
As we shall see, this difference can be accounted for within a refinement of the P model in 
different ways.  In real proteins, side chains are typically directed parallel to the outward 
normal \cite{Park96,Banavar06}. Upon folding, they have to be positioned in a way to avoid 
steric overlap, a condition met by both alpha-helices and beta-sheets in real proteins. 
One possible refinement of a polymer model therefore hinges on the inclusion of additional 
spherical beads that would mimick this effect. We call this model Polymer with Side Chains 
(PSC) model. The presence of side chains in the direction of the outward normal from the 
backbone additionally breaks the spherical symmetry of the original P model in favour of a 
cylindrical symmetry. This can be accounted for by allowing partial interpenetration of 
consecutive spherical backbone beads. This model will be referred to as Overlapping Polymer 
(OP), and will be the third model that we will consider in addition to the P and the PSC models. 
A final possibility is clearly given by combining the presence of side chains and the overlapping 
backbone spheres. This ultimate model will be denoted as Overlapping Polymer with Side Chains 
(OPSC) model. 

In real proteins, all these effects drastically reduce the huge degeneracy of the 
polymer collapsed state, by progressively removing the corresponding glassy nature of the original 
free energy landscape  \cite{Finkelstein02}. As a result, one eventually obtains a unique native 
state, rather than a multitude of local minima having comparable energies. Variants of the P model 
have also been used in the framework of Go-like models \cite{Taketomi75,Clementi00,Koga00,Badasyan08} 
routinely adopted in protein folding studies, where the amino acid specificities are enforced by including 
the native contact list into the simulation scheme. 

One of the main difficulties involved in numerical simulations of long polymer chains, stems from the very 
large computational effort necessary to investigate its equilibrium properties. This is true both when using 
conventional canonical techniques \cite{Allen87,Frenkel02}, as well as for more recently developed micro-canonical 
approaches, such as the Wang-Landau method \cite{Wang01}. Even in the simple P model, while high temperature 
behavior poses little difficulties, low-temperature/low-energy regions are much more problematic. With the 
canonical ensemble simulations the system frequently gets trapped into metastable states at low temperatures, 
and with the Wang-Landau method the low temperature results strongly depend on the employed low energy cut-off, 
as lower values require increasingly larger computational efforts.  It is then of paramount importance to discuss 
a correct implementation of such approaches within the framework of protein-like system and to make a critical 
assessment of the reliability of the corresponding results.

The present paper presents a first attempt of a systematic approach in the framework of the four proposed models 
(P,OP,PSC,OPSC). Using Wang-Landau micro-canonical technique \cite{Wang01}, supported by replica exchange canonical 
Monte Carlo simulations \cite{Frenkel02}, we discuss results stemming from the four models that provide a link between 
conventional synthetic polymers and protein-like systems. In the P model, the comparative  analysis with previous 
results allows us to identify the optimal conditions of applicability of the method. We then proceed to study a step 
by step generalization introduced by the OP, PSC and the combined OPSC models, using Wang-Landau method and comparing 
with previous results, when available, obtained using the replica exchange canonical approach. In all cases, we obtain 
the complete phase diagram in the temperature-interaction range plane. Interestingly, the phase diagrams in the relevant 
region show  a single coil-helix or a double coil-hybrid-helix transitions, in close analogy with the single coil-crystal 
or double coil-globule-crystal transitions in the polymer model.

As a by-product of our analysis, we re-obtain some of the results of Ref. \cite{Banavar09}, with particular emphasis 
on the role of the interaction range that is argued to be crucial for the phase diagram. In all the above cases, we 
use condition realistic for protein systems.

The paper is organized as follows. Section \ref{sec:model} describes the four different models used in the present 
study, followed by Section \ref{sec:thermodynamics} describing the connection to thermodynamics in both the 
microcanonical and canonical representations. Section \ref{sec:MC} illustrates the computational Monte Carlo 
techniques, and Section \ref{sec:results} the corresponding obtained results. Section \ref{sec:correlations} addresses 
the onset of the Fisher-Widom line in our system. The paper will be closed by Section \ref{sec:conclusions} reporting 
some concluding remarks and perspectives.
%%%%%%%%%%%%%%%%%%%%%%%%%%%%%%%%%%%%%%%%%%%%%%%%%%%%%%%%%%%%%%%%%%%%%%%%%%%%%%%
\section{From polymers to proteins: four different models}
\label{sec:model}
\begin{figure}
\includegraphics[width=1.0\columnwidth]{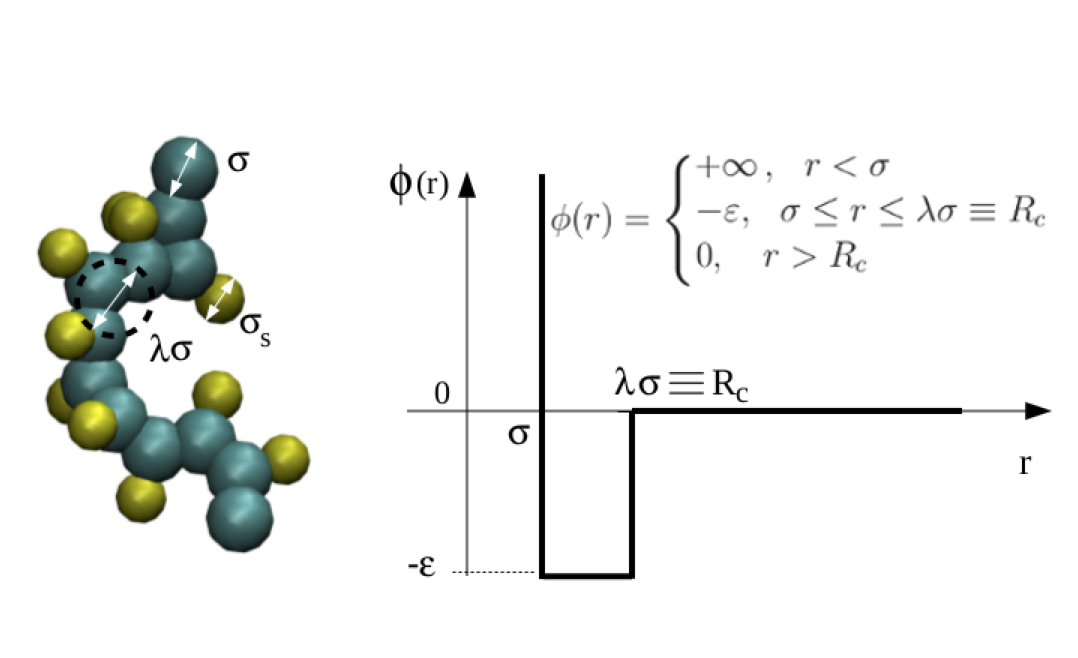}
\caption{The square-well potential used in the models. $\sigma$ is the diameter of the beads in the 
backbone, $\sigma_s$ that of the beads representing the side chains, $-\epsilon$ is the depth of the well.}
\label{fig:fig1}
\end{figure}

%%%%%%%%%%%%%%%%%%%%%%%%%%%%%%%%%%%%%%%%%%%%%%%%%%%%%%%%%%%%%%%%%%%%%%%%%%%%%
Following the standard approach \cite{Taylor09_a,Taylor09_b}, we model a polymer as a flexible chain formed by 
a sequence of $N$ monomers, located at positions $\{\mathbf{r}_1,\ldots,\mathbf{r}_N\}$, each having diameter 
$\sigma$. Consecutive monomers are connected by a tethering potential keeping the $N-1$ consecutive monomers 
at a fixed bond length equal to $b$. Non-consecutive monomers interact {\sl via} a square-well (SW) potential
\begin{equation}
\phi\left(r\right)=\begin{cases}
+\infty\,,\quad \,r < \sigma &\\
-\epsilon,\quad\, \sigma < r < R_c\equiv \lambda \sigma&\\
0, \qquad r > \lambda \sigma&
\end{cases}
\label{sw0}
\end{equation}
\noindent where $r_{ij}=|\mathbf{r}_{ij}|=|\mathbf{r}_j-\mathbf{r}_i|$, and $\lambda-1$ is the well width in units 
of $\sigma$ and defines the range of interaction $R_c=\lambda \sigma$.  Here $\epsilon$ defines the well depth 
and thus sets the energy scale (see Fig.\ref{fig:fig1}). The model has a discrete spectrum given by 
$E_n=-\epsilon n$, where $n$ is the number of SW overlaps \cite{Taylor09_b} that, in turn, depends upon $\lambda$.

In addition to the $N$ backbone beads, we will also consider $(N-2)$ side chain beads of hardcore diameter $\sigma_s$.
For each of non-terminal backbone beads, one defines a tangent and a normal vector as
\begin{eqnarray}
\widehat{\mathbf{T}}_i & = & \frac{\mathbf{r}_{i+1} - \mathbf{r}_{i-1}}
{|\mathbf{r}_{i+1} - \mathbf{r}_{i-1}|} \\
\widehat{\mathbf{N}}_i & = & \frac{\mathbf{r}_{i+1}-2\mathbf{r}_i+\mathbf{r}_{i-1}}
{|\mathbf{r}_{i+1}-2\mathbf{r}_i+\mathbf{r}_{i-1}|}
\end{eqnarray}
where $i=2,\ldots,N-1$. Consequently, one can also define a binormal vector as:
\begin{equation}
\widehat{\mathbf{B}}_i = \widehat{\mathbf{T}}_i \times \widehat{\mathbf{N}}_i .
\end{equation}
Note that $\{\widehat{\mathbf{B}}_i,\widehat{\mathbf{T}}_i,\widehat{\mathbf{N}}_i\}$, ($i=2,\ldots,N-1$) are the 
discretized version of the Frenet-Serret local coordinate frame description \cite{Coexter89} that is frequently 
used in the continuum approach of the polymer theory \cite{Kamien02,Poletto08,Bhattacharjee13}.

As we shall see, particularly useful in identifying the conformal structure of the resulting phases, are the 
tangent-tangent correlation function
\begin{eqnarray}
  \label{eq:tangent-tangent}
  G_l&=& \left \langle \widehat{\mathbf{T}}_i \cdot \widehat{\mathbf{T}}_{i+l} \right \rangle
\end{eqnarray}
and the average triple scalar product 
$\langle \widehat{\mathbf{N}}_i \cdot (\widehat{\mathbf{N}}_j \times \widehat{\mathbf{N}}_k) \rangle $ 
between any triplet of normal vectors, the meaning of the averages being defined in the next section.
The tangent-tangent correlation function drops exponentially for an unstructured coil chain configuration, 
while it becomes oscillating for a helical structure and it will be then used to identify helices.
On the other hand, the vanishing of the triple scalar product will be used to highlight the presence of a 
planar structure.

To each non-terminal backbone bead a side chain bead is attached in the anti-normal direction with the positions
given by:
\begin{equation}
\mathbf{r}^{(\text{s})}_i = \mathbf{r}_i - \widehat{\mathbf{N}}_i (\sigma + \sigma_s)/2 .
\end{equation}
The potentials involving side chain beads are just hardcore repulsions that vanish when $\sigma_s \to 0$.

\begin{figure}
\includegraphics[width=0.85\columnwidth]{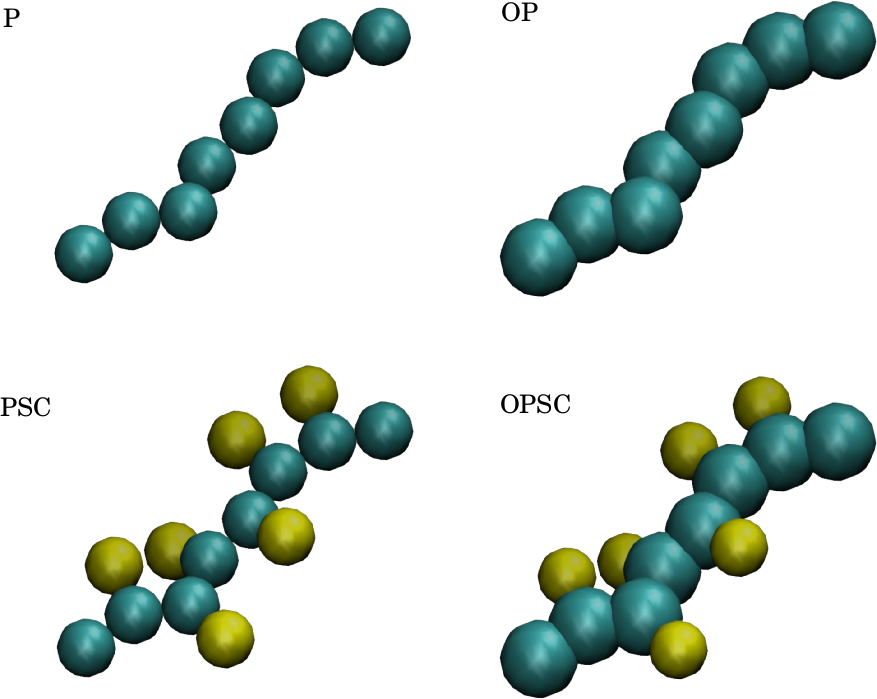}
\caption{The four different models considered in the present study. (P) has tangent backbone beads $b/\sigma=1$ 
and no side chains $\sigma_s = 0$. (OP) has no side chains but its backbone is formed by overlapping beads  $b/\sigma<1$. 
(PSC) has side chains  $\sigma_s > 0$ but no overlap $b/\sigma=1$. (OPSC) has side chains $\sigma_s > 0$ and overlapping 
beads in the backbone $b/\sigma<1$.}
\label{fig:fig2}
\end{figure}

We will then consider four different cases, schematically illustrated in Fig. \ref{fig:fig2}. In the simplest 
case of the Polymer model, denoted as (P) in Fig. \ref{fig:fig2}, $b/\sigma=1$ and there are no side chains 
($\sigma_s = 0$). In the Overlapping Polymer ((OP) model in Fig. \ref{fig:fig2}), consecutive backbone beads 
are allowed to overlap ($b/\sigma<1$) but still there are no side chains ($\sigma_s = 0$). The third intermediate case, 
that we dubbed Polymer with Side Chains (PSC) model, $b/\sigma=1$ but side chains are included ($\sigma_s > 0$).  
Finally, in the Overlapping Polymer with Side Chains (OPSC) model, backbone beads are allowed to overlap ($b/\sigma<1$) 
and beads representing side chains are present ($\sigma_s > 0$).

Henceforth, we will assume $\sigma$ and $\epsilon$ as the units of lengths and energies respectively, and investigate 
different ratios $b/\sigma$, side chain bead sizes $\sigma_s/\sigma$, as well as interaction ranges 
$\lambda=R_c/\sigma$. Eventually, connections with realistic values in the case of proteins will be established.

%%%%%%%%%%%% Fig 1  %%%%%%%%%%%%%%%%%%%%%%%%%%%%%%%%%%%%%%%%%%%%%%%%%%%%%%%%%%%%%%%%%%%%%%%%%%%%%%%%%%%%%%%%%%%%%%%%%%
%\begin{figure}[ht]
%%%%%%%%%%%%%%%%%%%%%%%%%%%%%%%%%%%%%%%%%%%%%%%%%%%%%%%%%%%%%%%%%%%%%%%%%%%%%%%%%%%%%%%%%%%%%%%%%%%%%%%%%%%%%%%%%%%%%%%%
%%%%%%%%%%%% Fig 2  %%%%%%%%%%%%%%%%
%\begin{figure}[ht]
%%%%%%%%%%%%%%%%%%%%%%%%%%%%%%%%%%%
%%%%%%%%%%%%%%%%%%%%%%%%%%%%%%%%%%%%%%%%%%%%%%%%%%%%%%%%%%%%%%%%%%%%%%%%%%%%%%%
\section{Thermodynamics in the Wang-Landau approach}
\label{sec:thermodynamics}
%%%%%%%%%%%%%%%%%%%%%%%%%%%%%%%%%%%%%%%%%%%%%%%%%%%%%%%%%%%%%%%%%%%%%%%%%%%%%
In the micro-canonical approach, the central role is played by the density of states (DOS) $g(E)$ that is related 
to the micro-canonical entropy as
\begin{eqnarray}
\label{entropy}
S\left(E\right) &=& k_B \ln g\left(E\right),
\end{eqnarray}
($k_B$ is the Boltzmann constant) and hence to the whole thermodynamics.
Canonical averages can be also computed using the partition function
\begin{eqnarray}
Z\left(T\right)&=&\sum_{E} g\left(E\right)e^{-E/\left(k_B T\right)},
\label{Z}
\end{eqnarray}
and the probability function
\begin{eqnarray}
P\left(E,T\right)&=&\frac{1}{Z\left(T\right)}g\left(E\right)e^{-E/\left(k_B T\right)},
\label{Prob}
\end{eqnarray}
to find the system in a conformation with the energy $E$ and temperature $T$. Helmholtz and internal energies can 
then be obtained as
\begin{eqnarray}
F\left(T\right)=-k_B T\ln Z\left(T\right) \\
U\left(T\right)=\left \langle E \right \rangle=\sum_{E} E P\left(E,T\right) 
\label{averages}
\end{eqnarray}
where the average $\langle \ldots \rangle$ is over the probability (\ref{Prob}).

As we shall see, two important probes of the properties of polymers and the transitions between different ordered 
states are given by the heat capacity
\begin{eqnarray}
C\left(T\right)&=&\frac{dU\left(T\right)}{dT}=
\frac{\left \langle E^2 \right \rangle-\left \langle E\right \rangle^2}{k_B T^2}
\label{Cv}
\end{eqnarray}
and by an averaged value of the radius of gyration \cite{Barrat03} $R_g^2=\sum_{i<j}^N (\mathbf{r}_i-\mathbf{r}_{j})^2/N^2$. This can obtained in microcanonical ensemble by constructing the distribution  $P(Rg^2,E)$ of $R_g$ at a given $E$
to get the micro-canonical average  $\langle \ldots \rangle_E$ over this distribution
\begin{eqnarray}
\left \langle R_g^2 \right \rangle_{E}&=&\sum_{R_g} P\left(R_g^2,E\right) R_g^2
\label{RgE}
\end{eqnarray}
as well as in the canonical ensemble using
\begin{eqnarray}
\left \langle R_g^2(T) \right \rangle &=&\sum_{E} \left \langle R_g^2\right \rangle_{E} g\left(E\right) 
e^{-E/\left(k_B T\right)}.
\label{RgT}
\end{eqnarray}
It is important to remark that the derivation of thermodynamics from the $g(E)$ is quite general, and does not 
depend on the specific method of simulation.
%%%%%%%%%%%%%%%%%%%%%%%%%%%%%%%%%%%%%%%%%%%%
\section{Monte Carlo simulations}
\label{sec:MC}
%%%%%%%%%%%%%%%%%%%%%%%%%%%%%%%%%%%%%%%%%%%%%
In this section, we briefly recall the main points of the Wang-Landau technique as applied to a single polymer chain, 
as well as the main differences with its canonical counterpart.
%%%%%%%%%%%%%%%%%%%%%%%%%%%%%%%%%%%%%%%%%%%%%%%%%%%%%%%%%%%%%%%%%%%%%%%%%%%%%%%
\subsection{Micro-canonical approach: Wang-Landau method}
\label{subsec:micro}
%%%%%%%%%%%%%%%%%%%%%%%%%%%%%%%%%%%%%%%%%%%%%%%%%%%%%%%%
Following general established computational protocols \cite{Allen87,Frenkel02}, and Refs.~\cite{Taylor09_a,Taylor09_b} 
for the specificities related to the polymers, we use Wang-Landau (WL) method \cite{Wang01} to sample polymer 
conformations according to micro-canonical distribution, by generating a sequence of chain conformations $a \to b$, 
and accepting new configuration $b$ with the micro-canonical acceptance probability
\begin{equation}
P_{acc}(a \rightarrow b)=\min{ \left(1,\frac{w_b g(E_a)}{w_a g(E_b)} \right)},
\label{prob}
\end{equation}
\noindent where $w_a$ and $w_b$ are weight factors ensuring the microscopic reversibility of the moves.

A sequence of chain conformations is generated using a set of Monte Carlo moves, which are accepted or rejected 
according to Eq.~\ref{prob}. The set of Monte Carlo moves employed consists of local-type moves, such as single-bead 
crankshaft, reptation  and end-point moves; as well as of non-local-type moves such as pivot, bond-bridging and 
back-bite moves. All moves are chosen at random and one sweep of the Wang-Landau algorithm is considered completed 
as soon as at least N monomer beads have been displaced.

The bond-bridging moves and back-bite types of the moves are found to be important in order to sample correctly 
and efficiently the low-energy states. Both of them consist of selecting one chain end (bead $1$ or $N$, old end 
bead in the following) and attempting to connect it to an interior bead $i$ randomly chosen among its neighbourhood 
within $2\sigma$ range. In the bond-bridging version of the move \cite{Taylor09_b,Dellago14} this attempt is achieved
by removal of the bead next to the chosen one in the direction of the chosen end (bead $i-1$ or $i+1$) and its 
re-insertion between the bead $i$ and the old-end bead via a crankshaft type of the move. The factor that in this 
case ensures the microscopic reversibility of the move reads $w_b/w_a = n_a \mathrm{\,} r_b/n_b \mathrm{\,} r_a$, 
where $n_a$ and $n_b$ are the numbers of neighbours of the old and the new chain end within the $2\sigma$ range, 
respectively; $r_a$ and $r_b$ are the distances between (old and new) chain ends and the $i-$th bead.
In the back-bite version of the move \cite{Virnau10} the old chain end ($1$ or $N$) is attempted to be directly 
connected with the randomly selected bead $i$ in its neighbourhood and the correction of the acceptance probability 
in this case is simply $w_b/w_a = n_a /n_b$. The weight factors for other types of Monte Carlo moves are all equal to unity.
%%%%%%%%%%%%%%%%%%%%%%%%%%%%%%%%%%%%%%%%%%%%%%%%%%%%%%%%%%%
%%%%%%%%%%%%%%%%%%%%%%%%%%%%%%%%%%%%%%%%%%%%%%%%%%%%%%%%%%%
\subsection{Canonical approach: replica exchange}
\label{subsec:canonical}
%%%%%%%%%%%%%%%%%%%%%%%%%%%%%%%%%%%%%%%%%%%%%%%%%%%%%%%%%%%
The parallel tempering (or replica exchange) technique \cite{Swendsen86,Geyer91} is a powerful method for sampling 
in systems with rugged energy landscape. It allows the system to rapidly equilibrate and artificially cross energy 
barriers at low temperatures. Furthermore, the method can be easily implemented on a parallel computer. Parallel 
tempering technique can be used with both Monte Carlo and Molecular Dynamics simulations, but in this study, we apply 
this technique specifically to Monte Carlo simulation. The method entails monitoring $M$ canonical simulations in 
parallel at $M$ different temperatures,  $T_i$, $i=1,2,\ldots,M$. Each simulation corresponds to a replica, or a copy 
of the system in thermal equilibrium. In individual Monte Carlo simulations, new moves are accepted with standard 
acceptance probabilities given by the Metropolis method \cite{Metropolis}. The replica exchange technique allows 
the swapping of replicas at different temperatures without affecting the equilibrium condition at each temperature.  
Specifically, given two replicas $\Gamma_i$ being at $T_i$ and $\Gamma_j$ being at $T_j$, the swap move leads to a 
new state, in which $\Gamma_i$ is at $T_j$  and $\Gamma_j$ is at $T_i$. The acceptance probability of such a move can 
be derived based on the detailed balance and is given by:
\begin{eqnarray}
P_\mathrm{swap} &=& \min \left(1,\exp \left[ \left(\frac{1}{k_B T_i} - 
\frac{1}{k_B T_j}\right)(E_i - E_j)\right] \right).
\end{eqnarray}
The choice of replicas to perform an exchange can be arbitrary, but for a pair of temperatures, for which replicas 
are exchanged, the number of swap move trails must be large enough to ensure good statistics. The efficiency of a 
parallel tempering scheme depends on the number of replicas, the set of temperatures to run the simulations, how 
frequent the swap moves are attempted, and is still a matter of debate. It has been suggested that for the best 
performance, the acceptance rate of swap moves must be about 20\% \cite{Rathore05}.
%%%%%%%%%%%%%%%%%%%%%%%%%%%%%%%%%%%%%%%%%%%%%%%%%
\section{Results}
\label{sec:results}
%%%%%%%%%%%%%%%%%%%%%%%%%%%%%%%%%%%%%%%%%%%%%%%%%
%%%%%%%%%%%%%%%%%%%%%%%%%%%%%%%%%%%%%%%%%%%%%%%%%%%%%%%
\subsection{Model P}
\label{subsec:modelP}
%%%%%%%%%%%%%%%%%%%%%%%%%%%%%%%%%%%%%%%%%%%%%%%%%
Before tackling more complex systems, we will test our WL and replica exchange approaches using the simple, and yet 
interesting, P model ($\sigma/b=1$ and $\sigma_s=0$) for which several studies \cite{Taylor09_a,Taylor09_b} have been 
previously performed and can then be contrasted with.

The aim of the calculation is the computation of the DOS $g(E)$. In the WL method, $g(E)$ is constructed iteratively,  
with smaller scale refinements made at each level of iteration, controlled by the flatness of energy histogram. We typically consider an iteration  to reach convergence after $30$ levels of iteration, corresponding to a multiplicative factor values of $f=10^{-9}$.  This choice is neither unique, nor universally accepted \cite{Seaton09,Zhou05,Belardinelli07,Swetnam11}, but we have checked this to be sufficient for our purposes by comparing results obtained with different values. An additional crucial step in WL algorithm hinges in the selection of ground state energy. As this must be defined at the outset, and is known to drastically affect the low-energy behavior of the system \cite{Wust08,Seaton09}, care must be exercised in its selection. At the present time, however, there is no universally accepted procedure for off-lattice Wang-Landau simulations. Here we will be following the procedure suggested in Refs. \cite{Taylor09_a,Taylor09_b}, that has been reported to be reliable in most cases. A preliminary run with no low-energy cutoff is carried out for a sufficient number of MC steps ($10^9$ in our case). This provides an estimate of the low energy cut-off, that can then be increased by $1\%$ percent and used throughout successive calculations.

In our parallel tempering scheme we consider 20 replicas and the temperatures are chosen such that they decrease geometrically $T_{i+1}=\alpha T_i$,  where $\alpha=0.8$, starting from the highest reduced temperature $k_B T_1/\epsilon=10$, at which the polymer is well poised in the swollen phase. We allow replica exchange only  between neighboring temperatures and for each replica a swap move is attempted every 50 Monte Carlo steps. Standard pivot and crank-shaft move sets \cite{Sokal} are used  in Monte Carlo simulations. A typical length of the simulations is $10^9$ steps per replica.  Results from parallel tempering simulations are the equilibrium data and are convenient for analysis using the weighted histogram analysis method  \cite{Ferrenberg89}. The latter allows an estimate of the density of states as well as to calculate the thermodynamic averages from simulation data
at various equilibrium conditions.

\begin{figure}
\includegraphics[width=0.85\columnwidth]{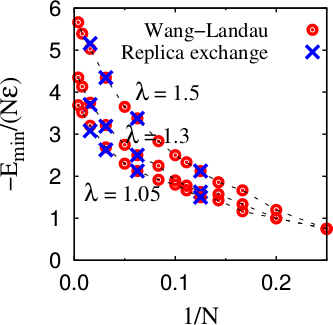}
\caption{Plot of the reduced minimum energy per monomer $-E_{min}/(N \epsilon)$ as a function of $1/N$ for 
$\lambda=1.5$, using both canonical (crosses) and Wang-Landau (circles) approaches.}
\label{fig:fig3}
\end{figure}

We have explicitly performed both procedures for chains ranging from $N=4$ to $N=128$, as depicted in Fig. \ref{fig:fig3} where the reduced minimum energy per monomer $-E_{min}/(N \epsilon)$ is plotted against $1/N$. The results are  in perfect agreement with each other and with those obtained
in the extensive simulations by Taylor \textit{et al}  \cite{Taylor09_a,Taylor09_b}.

%%%%%%%%%%%%%%%%%%% Fig 3 %%%%%%%%%%%%%%%%%%%%%%%%%%%%%%%%
%\begin{figure}[ht]
%%%%%%%%%%%%%%%%%%%%%%%%%%%%%%%%%%%%%%%%%%%%%%%%%%%%%%%
As an additional preliminary step, we have tested our WL code against exact analytical results valid for small 
$N=4$ (tetramers) and $N=5$ (pentamers) \cite{Taylor95,Taylor03,Magee08} at different values of the interaction range 
$\lambda$. A final test stems from a comparison with the Molecular Dynamics (MD) and canonical Monte Carlo (MC) results 
by Zhou and Karplus \cite{Zhou97}. In all cases, we have found complete agreement, as detailed in the supplementary 
information.

The successfull comparison with previous results of past literature on the P model, gave us confidence about the 
reliability of our computational tools. A particularly interesting results, originally derived  by Taylor \textit{et al}  
\cite{Taylor09_a,Taylor09_b} and reproduced by our calculations (see supplementary information Fig2S), hinges upon the
interaction range dependence of the low temperature state of the polymer. Upon cooling from a high temperature coil state, 
the chain P first exhibits a transition to a collapsed globule and then to a frozen crystallyte. However, for very 
short-range interaction ($\lambda < 1.05$), the continuous collapse transition is pre-empted by the freezing transition, 
and there is a direct first-order coil-crystallite transition. Interestingly, this range dependence mirrors a similar 
dependence destabilizing the liquid phase in favour of the solid phase in SW fluids \cite{Pagan05}.

As we will see below, we will observe something similar in the coil-helix transition of the OPSC model.
%%%%%%%%%%%%%%%%%%%%%%%%%%%%%%%%%%%%%%%%%%%%%%%%%
\subsection{From Model P to Model OPSC}
\label{subsec:modelP_to_modelOPSC}
%%%%%%%%%%%%%%%%%%%%%%%%%%%%%%%%%%%%%%%%%%%%%%%%%
Having discussed the P model as a suitable model for polymers, we now add additional ingredients in our strive toward 
a model for a protein. While the P model is perfectly able to describe the correct low temperature behaviour of polymers, 
it cannot be used for proteins not even at the minimal level of description. There are a number of reasons for that. 
Previous results \cite{Taylor09_a, Taylor09_b} briefly recalled in Section \ref{subsec:modelP}, unambiguously showed 
that a model of a chain formed by tethered spherical beads can only have a high temperature coil phase, an intermediate 
globular phase -- provided the range of interactions to be sufficiently long --  and a low temperature crystal phase. 
Due to its inherent isotropy, the P model cannot capture the essential ingredients for the formation of secondary 
structures in proteins.

One possibility to cope with this is to replace beads with disks, thus breaking the spherical symmetry. In a continuum 
description, this gives rise to the so-called tube (or thick polymer) model \cite{Maritan00,Banavar03}. This effect alone 
was shown to give rise to formation of helices and planar conformations \cite{Poletto08}. The tube model, however, 
requires a three-body potential \cite{Maritan00,Banavar03}, and is then very demanding from the computational viewpoint.
A similar effect can be achieved by allowing partial overlapping of consecutive spherical monomers, giving rise to model OP considered here. In the OP model consecutive spheres overlap ($b/\sigma<1$) but side chains are absent ($\sigma_s=0$). Another ingredient that is known to play a fundamental role in the low temperatures behavior of proteins is the steric hindrance of the side chains. This can be seen, for instance, from the Ramachandran plots in real proteins showing that both alpha helices and beta sheets are constrained into well defined regions of the dihedral angles \cite{Finkelstein02}. One can then envisage the possibility of introducing additional spherical beads, attached to each backbone bead at the appropriate location and representing the van der Waals spheres of the side chains. This leads to the PSC model that is complementary to the OP model in the sense that consecutive spheres do not overlap ($b/\sigma=1$) but side chains are present ($\sigma_s>0$). Finally, one can consider both effects, thus obtaining the OPSC model. Note that both backbone and side chain spheres are considered aspecific, unlike in real proteins where both have  their own specific character. The effect of this ingredient has been considered in the past and therefore will be neglected for simplicity in the present work, although in principle it could be accounted for without any difficulties.
%%%%%%%%%%%%%%%%%%%%%%%%%%%%%%%%%%%%%%%%%%%%%%%%
\subsection{Model OPSC }
\label{subsec:modelOPSC}
%%%%%%%%%%%%%%%%%%%%%%%%%%%%%%%%%%%%%%%%%%%%%%%%%
We now focus on the OPSC model as that of main interest for the present work.
The OPSC model was already considered in Ref. \cite{Banavar09} using the replica-exchange canonical simulations, both for short ($N=16$) and long ($N=50$) chains. The necessary effort for reaching the correct ground state was considerable for the longer chain case, due to the appearance of an increasing number of metastable configurations that might hamper the correct low temperature sampling. In the present study, we will extend results obtained in Ref.\cite{Banavar09} via a Wang-Landau method that will shed new light on the OPSC model within a wider perspective. We will obtain the complete phase diagram in the temperature-interaction range plane that highlights the fundamental role of the range of interaction on the formation of the secondary structures.

\begin{figure}
\includegraphics[width=0.85\columnwidth]{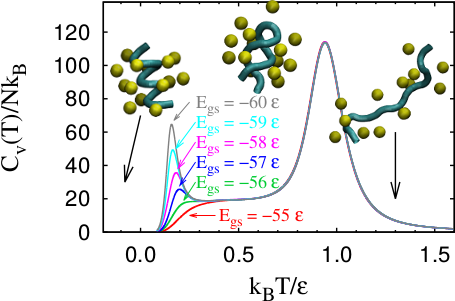}
\caption{Specific heat per monomer  $C_{V}/(N k_B)$ as a function of the reduced temperature $k_B T/\epsilon$ in the case $N=16$ with different ground state energies $-60 \le E_{gs}/\epsilon\le -55$ These results for $E_{gs}/\epsilon=-60$ are compatible with Fig.3a of Ref.\cite{Banavar09}.
The insets include snapshots of rapresentative configurations
(helix on the left, double helix in the middle and the coil on the right). Here and below, a tube representation of the backbone is employed for a better visual reproduction.
}
\label{fig:fig4}
\end{figure}

In OPSC consecutive spheres overlap ($b/\sigma<1$) and side chains are present ($\sigma_s>0$). Here, we will use both dimensionless and dimensional units to make contacts with both general features
of the P model previously discussed, and with realistic values characteristic of protein systems.
As in Ref.\cite{Banavar09}, we will use $N=16,50$, $\sigma=6${\AA} and $\sigma_s=5.0${\AA} and $b=3.8${\AA}. The range of interactions here is set to
$R_c=7.5${\AA}.
This corresponds to $b/\sigma \approx 0.63$, $\sigma_s/\sigma \approx 0.83$, and $\lambda=R_c/\sigma \approx 1.25$ in dimensionless units.
The WL calculation was performed by using different ground state energies from $E_{gs}/\epsilon=-55$ to $E_{gs}/\epsilon=-60$, and Fig. \ref{fig:fig4} reports the specific heat per monomer $C_{V}/(N k_B)$.
In this case, the lower bound energies are simply
set by the interval of values one is considering when filling the energy histograms. In practice, all energies lower
than these predefined cut-off are discarded.
Here $N=16$ as in Ref. \cite{Banavar09}. While the first peak at higher temperatures, associated with the coil-to-globule transition,   is unaffected by the actual value of the ground state, the different behavior at  low temperatures is clearly visible. When  $E_{gs}/\epsilon=-55$ the low temperature second peak is absent, indicating that the ground state is the double helix structure. Note that the double helix is stabilized by consecutive non-local contacts, in the same way as protein's $\beta$-sheets, but lacks the planar symmetry of the latter. As the energy decreases toward the correct ground state $E_{gs}/\epsilon=-60$, a second low temperatures peak emerges indicating an additional structural transition to a helical state, that is then the ground state of the system under these conditions.
Note that the results for $E_{gs}/\epsilon=-60$ are in full agreement with those obtained in Fig.3a of Ref.\cite{Banavar09} {\sl via} replica-exchange canonical simulations.
%%%%%%%%%%%%%%%%%%%%% Fig 4 %%%%%%%%%%%%%%%%%%%%
%\begin{figure}[ht]
%%%%%%%%%%%%%%%%%%%%%%%%%%%%%%%%%%%%%%%%%%%%%%%%

\begin{figure}
\includegraphics[width=0.85\columnwidth]{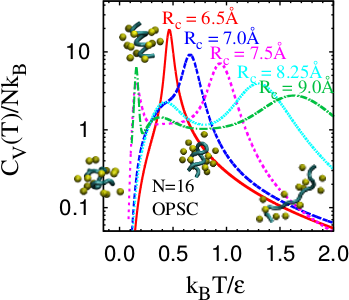}\\
\vspace{5mm}
\includegraphics[width=0.85\columnwidth]{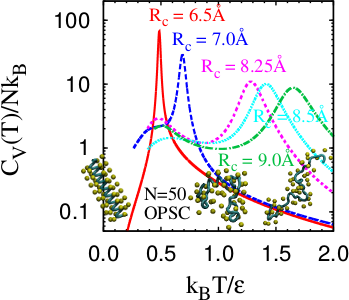}
\caption{(Top) Specific heat per monomer  $C_{V}/(N k_B)$ as a function of the reduced temperature $k_B T/\epsilon$ in the case $N=16$ with different interaction ranges $R_c$.
(Bottom) Same as above in the case $N=50$. In this case, the results for $R_c=6.5${\AA} are compatible with Fig.S2(a)in the supplementary information of Ref.\cite{Banavar09}.}
\label{fig:fig5}
\end{figure}

Even when the actual low energy limit is properly accounted for, the corresponding ground state is strongly dependent upon the chosen range of the interactions $R_c$. This is not surprising on physical grounds. As $R_c/\sigma \approx 1$, the range of attraction is essent

ers, and the chain will have the tendency to form a coil configuration at high temperatures and a suitable secondary structure maximizing  the number of favorable contacts at low temperatures. In the opposite limit $R_c/\sigma \gg 1$, each monomer can interact with essentially any other monomer in the chain, and at low temperature will tend to form a globular structure that displays the lowest possible energy. On the basis of results given in Fig. \ref{fig:fig4} for $N=16$, one could reasonably expect a formation of a stable helix for some intermediate value  of the $R_c/\sigma$ ratio.  This is confirmed in Fig. \ref{fig:fig5}, where we report the specific heat per monomer $C_{V}/(N k_B)$ as a function of the reduced temperature
$k_B T/\epsilon$ in the case $N=16$ (top panel) and $N=50$ (bottom panel), at different ranges of interactions $R_c$. In the former case ($N=16$) a single well-developed peak is clearly visible for  $R_c=6.5${\AA}, corresponding to the coil-helix transition. As $R_c$ increases, this single peak gradually evolves first into two, around $R_c \approx 7.5${\AA}, and eventually into three at $R_c \approx 9.0${\AA}, accounting for an intermediate double helix configuration at  $k_B T/\epsilon \approx 0.8$, and for a globular configuration at  the lowest $k_B T/\epsilon$. The $N=16$ result is particularly useful because it allows the presence also of larger values of $R_c$ that typically require long computational time.

We note that these results extend those presented in Fig.S2(a) in the supplementary information of Ref.\cite{Banavar09} obtained by  replica-exchange canonical simulations. In all cases the ground state energies were found identical (dependent on the choice of $R_c$).

%%%%%%%%%%%%%%%%%%%%% Fig 5 %%%%%%%%%%%%%%%%%%%%
%\begin{figure}[ht]
%%%%%%%%%%%%%%%%%%%%%%%%%%%%%%%%%%%%%%%%%%%%%%%%
The case $N=50$ allows for a better statistics at the expense of an increased computational effort,
see  Fig. \ref{fig:fig5} (bottom panel). A striking feature of Fig. \ref{fig:fig5} is again the
presence of a single strong peak indicating the ground state of the system corresponding to a helical
organization, as illustrated in the insert snapshot, along with a secondary peak at slightly higher
temperatures. The second peak increases, and in fact becomes the dominant one, upon increasing the
interaction range $R_c$, indicating the progressive stabilization of the double helix
phase found at intermediate temperatures. This is in line with previous results for $N=16$ with the proviso that the globular phase is missing here due to the fact that the computational effort becomes prohibitively large for $R_c \gg 9$ {\AA}, a region where this phase is expected. Compatible results (not shown) can be obtained also by considering the gyration radius, in analogy to what done in the P model case.

\begin{figure}
\includegraphics[width=0.85\columnwidth]{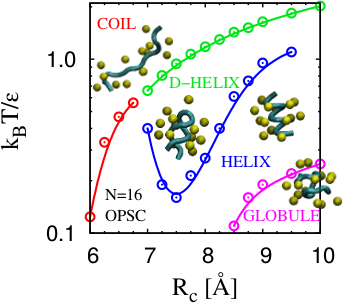}\\
\vspace{5mm}
\includegraphics[width=0.85\columnwidth]{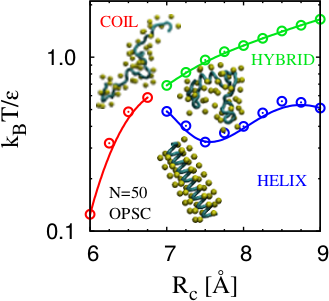}
\caption{Phase diagram of the OPSC model in the temperature-interaction range plane.(Top) $N=16$ case. (Bottom) $N=50$ case.
}
\label{fig:fig6}
\end{figure}

Guided by the above findings, we can now draw the phase diagram in the temperature $k_B T/\epsilon$ - interaction range $R_c$ plane. This complements those already obtained in Ref.\cite{Banavar09}
in the $\sigma_s-\sigma$ plane and is reported in Figure \ref{fig:fig6} in the case $N=16$ (top panel) and $N=50$ (bottom panel). Here $\sigma=6${\AA} and
$\sigma_s=5$ {\AA}.
%%%%%%%%%%%%%%%%%%%%% Fig 6 %%%%%%%%%%%%%%%%%%%%
%\begin{figure}[ht]
%%%%%%%%%%%%%%%%%%%%%%%%%%%%%%%%%%%%%%%%%%%%%%%%
Several interesting features are apparent. Starting from a coil high temperature configuration, one finds a direct
transition to a helix conformation upon cooling for sufficiently short interaction ranges ($R_c \le 7$ \AA). Above
this value, there is first a  transition to a double helix structure followed by a successive transition to helix
structure. The two transitions are expected to merge at large interaction range, where the low temperature ground
state is presumed to be a globule, irrespective of $R_c$.
Note that the characteristic dual helix structure appearing in the $N=16$ case
evolves into a more complex structure where the internal helical shape is bracketed by double-helicies at chain endings,
connected via flexible turns. We dubbed this hybrid configurations in the lower panel of Figure \ref{fig:fig6}.
%%%%%%%%%%%%%%%%%%%%%%%%%%%%%%%%%%%%%%%%%%%%%%%%%
\subsection{Model OP}
\label{subsec:modelOP}
%%%%%%%%%%%%%%%%%%%%%%%%%%%%%%%%%%%%%%%%%%%%%%%%%
We now backtrack to investigate the effect of removing one-by-one single ingredients in the resulting phase diagram.

In the OP model, consecutive beads overlap, as in the OPSC model, but here the side chains are missing ($\sigma_s=0$), so that the steric hindrance of the side chain beads is also absent. On the other hand, partial interpenetration of consecutive beads breaks the spherical symmetry of the P model, along the backbone direction. Effectively, this provides a restriction on the number of possible local conformations that the chain can achieve as its effective shape becomes more cylindrical. A similar effect can be obtained in the tube model by enforcing a finite thickness of the chain but it requires a three-body potential \cite{Maritan00}. In the $N=16$ case, for high and low temperatures the coil and globule conformations are respectively observed, with intermediate temperature phases typically as the double helical conformation at low $R_c$ and a helicoidal conformation at larger $R_c$. This agrees with findings from the tube model, in spite of the different driving forces in the two systems \cite{Poletto08}.
These results also complement those obtained via replica exchange simulations by Magee and collaborators \cite{Magee07,Magee08} who observed helix formation in the OP model for sufficiently high interpenetration with $N=20$ but did not compute the phase diagram, as done here.

\begin{figure}
\includegraphics[width=1.0\columnwidth]{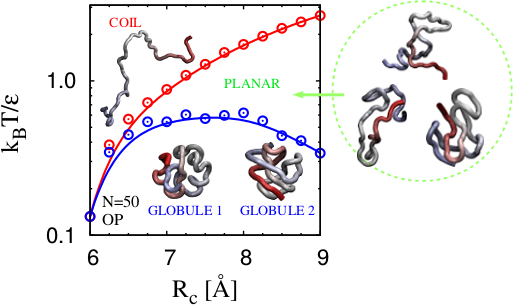}
\caption{Phase diagram of the OP model in the temperature-interaction range plane in the $N=50$ case.}
\label{fig:fig7}
\end{figure}

As the length of the polymer increases, we find these intermediate  double helical  and helicoidal configurations to become less and less stable, so that in the case of $N=50$ only a stable coil and globular configurations, with some intermediate planar-like structures, are observed. In Figure \ref{fig:fig7}, we display the obtained phase diagram in the case $N=50$. Interestingly, below $R_c \approx 6$ \AA, we expect a direct coil-globule transition.

A final remark is here in order. While it is intuitive that the OP model shares some similarities with the tube model discussed in Ref. \cite{Maritan00,Poletto08}, since both lead to a breaking  of the spherical symmetry with the appearance of a privileged axis along the chain backbone, we believe that some of the specificities of the three-body potential involved in the tube model cannot
be represented by a two-body potential as discussed here.
%%%%%%%%%%%%%%%%%%%%% Fig 7 %%%%%%%%%%%%%%%%%%%%
%\begin{figure}[ht]
%%%%%%%%%%%%%%%%%%%%%%%%%%%%%%%%%%%%%%%%%%%%%%%%
\subsection{Model PSC}
\label{subsec:modelPSC}
\begin{figure}
\includegraphics[width=0.75\columnwidth]{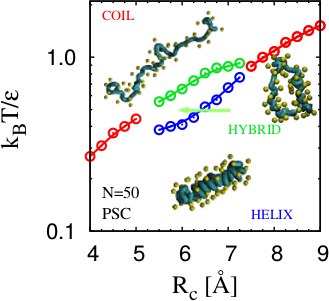}
\caption{Phase diagram of the PSC model in the reduced temperature $k_BT/\epsilon$ vs interaction range $R_c$ plane 
for $N=50$. Here $\sigma=b=3.8${\AA} and $\sigma_s=5$ {\AA}.}
\label{fig:fig8}
\end{figure}

%%%%%%%%%%%%%%%%%%%%%%%%%%%%%%%%%%%%%%%%%%%%%%%%%
The OPSC model reduces to the PSC model in the absence of overlapping ($b/\sigma=1$). This was already partially considered in Ref.\cite{Banavar09}, where again a phase diagram  in the $\sigma_s-\sigma$
plane was obtained. Our resulting phase diagram is displayed in Fig.\ref{fig:fig8}. When contrasted with the corresponding OPSC counterpart of Fig.\ref{fig:fig6}(bottom), we note a significant
shrinking of the intermediate phase, as well as a reduced shape stability of the helical phase.
All in all, our interpretation of these findings is that the interpenetrability of the backbone beads and the presence of the side chain beads are both crucial concurring ingredients  to
the observation of secondary structures, a combination of the two providing an optimal condition to their stability.
%%%%%%%%%%%%%%%%%%%%% Fig 8 %%%%%%%%%%%%%%%%%%%%
%\begin{figure}[ht]
%%%%%%%%%%%%%%%%%%%%%%%%%%%%%%%%%%%%%%%%%%%%%%%%
%%%%%%%%%%%%%%%%%%%%%%%%%%%%%%%%%%%%%%%%%%%%%%%%%
\section{Correlation function and the Fisher--Widom line}
\label{sec:correlations}
%%%%%%%%%%%%%%%%%%%%%%%%%%%%%%%%%%%%%%%%%%%%%%%%%

The Landau-Peierls theorem \cite{Landau80,Thouless69} forbids the presence of a thermodynamic phase transition in a strictly one-dimensional system with only short range interactions, and
a similar result holds true for liquid systems \cite{VanHove50}.

In fluid systems and in the presence of competing attractive/repulsive interactions, such as those occurring here, a Fisher--Widom (FW) line  (not to be confused with the Widom line \cite{Franzese07}) is separating two different regimes in the radial distribution function \cite{Fisher69}, characterized by an oscillatory (when repulsions are dominating) and an exponentially decaying (when attractions are dominating) behavior. The rationale behind the FW line is that on approaching the critical points where attractions become more and more effective, the behavior of correlation functions must switch from oscillatory (characteristic of repulsive interactions) to exponential with a well defined correlation length. The FW line was also observed in one-dimensional fluid of penetrable spheres \cite{Fantoni09,Fantoni10}, where Landau-Peierls theorem is no longer valid. In practice, the FW line is associated with an abrupt discontinuity in the structural behaviour of the system (as signalled by the correlation function) that is not, however, related to any real discontinuity in the thermodynamical behaviour.

As shown below, a similar feature occurs in the present framework where
the abrupt change in a correlation function is indicative of a structural transition, preceeding (i.e. occurring at slightly higher temperatures)
the actual thermodynamical transitions, that will then be interpreted as the countepart of the FW line.

\begin{figure}
\includegraphics[width=0.95\columnwidth]{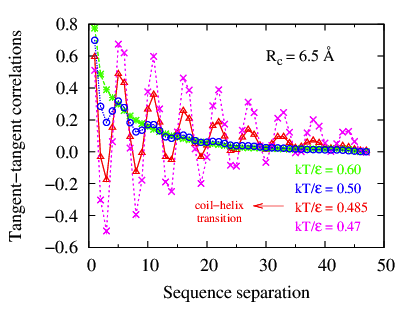}
\caption{Plot of the tangent-tangent correlation functions as a function of the sequence separation at four different 
reduced temperatures $k_B T/\epsilon$: above the Fisher-Widom line ($k_B T/\epsilon=0.6$), at the Fisher-Widom line 
($k_B T/\epsilon=0.5)$, at the coil-helix transition ($k_B T/\epsilon=0.485)$, and below it ($k_B T/\epsilon=0.47)$.}
\label{fig:fig9}
\end{figure}

Consider the tangent-tangent correlation function $G_l$ defined in Eq.(\ref{eq:tangent-tangent}) as a function of the sequence separation
(distance of the beads along the chain) $s$. In the coil state, the conformation of the chain is random and hence there is no correlation between the orientations of consecutive backbone beads. As a result, the tangent-tangent correlation is an exponentially decaying function of the sequence separation. In the helix state, however, the tangent-tangent correlation is an oscillating function of the sequence, with periodicity of about 4 units. The transition between the two regimes is an abrupt one, and occurs at temperatures slightly higher than those associated with the actual coil-to-helix transformation. Figure \ref{fig:fig9} confirms this expectation. Here the tangent-tangent correlation is plotted as a function of the sequence separation at four different reduced temperatures,  $k_B T/\epsilon=0.6$ above the FW line, $k_B T/\epsilon=0.5$ at the FW line, $k_B T/\epsilon=0.485$ at the coil-helix transition, and $k_B T/\epsilon=0.47$ below it. The transition from an exponential to an oscillating behavior upon cooling below the FW line is thus evident. Note that this transition can be experimentally probed by e.g. circular dichroism, and some straightforward  procedures have been devised to interpret the results  of these measurements in terms of simple spin models \cite{Badasyan10,Badasyan12}. Similar transitions also occur for the normal-normal and binormal-binormal correlations as shown in the supplementary information (see Figures 3Sa, 3Sb and 3Sc).

Our findings agree with those by Banavar \textit{et al} \cite{Banavar09} obtained {\sl via} the replica-exchange method, where the onset of oscillations in the tangent-tangent correlation as a function
of the sequence separation, was taken as an indicator of the incipient coil-helix transition.
%%%%%%%%%%%%%%%%%%%%% Fig9  %%%%%%%%%%%%%%%%%%%%
%\begin{figure}[ht]
%%%%%%%%%%%%%%%%%%%%%%%%%%%%%%%%%%%%%%%%%%%%%%%%
%%%%%%%%%%%%%%%%%%%%%%%%%%%%%%%%%%%%%%%%%%%%%%%%%%%%%%%%%%%%%%%%%%%%%%%%%%%%%%%%%%%%%%%%%%%%%%%%%%%%%%%%%%%%%%%%%%%%%%%%%%%%%%
\section{Conclusions}
\label{sec:conclusions}
%%%%%%%%%%%%%%%%%%%%%%%%%%%%%%%%%%%%%%%%%%%%%%%%%%%%%%%%%%%%%%%%%%%%%%%%%%%%%%%%%%%%%%%%%%%%%%%%%%%%%%%%%%%%%%%%%%%%%%%%%%%%%
In this paper we have studied the equilibrium statistics of a homopolymer formed by a sequence of tangent identical monomers represented by impenetrable hard spheres. A tethering potential keeps the consecutive backbone beads linked, whereas non-consecutive ones interact via a square-well potential, that then drives the collapse of the chain at sufficiently low temperatures. Three different variants of the above P model were considered as a gradual step toward a more realistic model for proteins. By allowing consecutive backbone spheres to interpenetrate, one breaks the rotational symmetry of the P model, thus opening up the possibility of having additional intermediate transitions. This model, dubbed the OP model, is similar in spirit to the tube model \cite{Maritan00,Banavar03}, where the symmetry breaking was enforced by replacing spherical with disk-like beads. Another variant, denoted as PSC model, keeps the backbone beads tangent, but allows the presence of additional side-chain beads located at appropriate position of the side chain center in real proteins. The interest in this variant is triggered by some some studies suggesting the importance of the role of side chains in the formation of secondary
structures \cite{Yasuda10}. The last variant, the OPSC model, combines all of the above effects.

As a preliminary validation of our approach, we first reproduced some of the Wang-Landau results obtained by Taylor \textit{et al} \cite{Taylor09_a,Taylor09_b} for model P, with long chains (with up to $N=256$ monomers) that we also re-obtained via replica-exchange canonical approach. This comparison allows us to be confident in the reliability of our computational scheme.

Having done that, we have then used the same Wang-Landau method to tackle OP, PSC, and OPSC models. The latter can be regarded as a minimal model for the formation of secondary structures in proteins and was studied by replica-exchange techniques in Ref. \cite{Banavar09}. In particular, we were able to underpin the fundamental role played by the interaction range in selecting the secondary structure. For very short range square-well attractions, the lowest energy is achieved by a double helix, in agreement with our intuition and with previous findings on the tube model \cite{Poletto08}. At intermediate interaction range, the lowest energy is associated with the formation of a well-defined helical structure, whereas at interaction ranges much longer of the bead sizes, the ground state is a globule, as expected in view of its mean-field like character. The possibility of tuning a single coil-to-globule or a double coil-to-helix and helix-to-globule transition, can be regarded as the  OPSC model counterpart of the results found in the P model by Taylor \textit{et al} \cite{Taylor09_a,Taylor09_b}, where the double transition coil-to-globule and globule-to-crystal transitions found at moderate attractive range, are replaced by a direct coil-to-crystal transition below a critical interaction range, in close analogy with what is found in square-well fluids, where a double
gas-liquid and liquid-solid transitions is preempted by a direct gas-solid transition below a critical range of interactions \cite{Pagan05}. Note, however, that this parallelism limits to an
analogy as the coil-helix transition is a structural rather than a thermodynamical transition.

In addition to to the OPSC model, we have also studied the simpler OP model, where side chains are absent, and the PSC model, where there is no overlap between consecutive spheres. We found that
both models display a phase transition similar to the full OPSC model but with a larger degree of metastability.

Finally, we have also discussed the transition in the behavior of the tangent-tangent correlation functions, as well as the
the normal-normal and the binormal-binormal correlation functions, as a function of the temperature. We found that the exponental behaviour in the sequence separation characteristic of the coil configuration at high temperatures, is replaced by an oscillatory behaviour at lower temperatures, indicating a helical organization, and this occurs at temperatures higher than the actual coil-helix thermodynamic transition. We proposed this line to be regarded as the analogue of the Fisher-Widom line found in fluid systems, in such this structural transition precedes the actual thermodynamical transition
associated with the discontinuity in the thermodynamical variables.

The formation of secondary structures is an important intermediate step for the general process of the folding of a protein. Our results are consistent with previous studies
\cite{Maritan00,Banavar03} suggesting that the introduction of a cylindrical (as opposed to spherical) symmetry is in fact sufficient to remove the characteristic glassy structure of the low energy state in polymers, and highlights the presence of side chains as a further stability factor. The rigidity of the chain, mostly ascribed to the presence of intrachain hydrogen bonds,
is known to be another important ingredient for protein folding that could further introduced in our analysis as a further step toward a more realistic model for a protein.
Work along these lines is currently underway and will be reported in a future study.

\appendix
\section{Phase diagrams of P model}
\begin{figure}
\includegraphics[width=0.73\columnwidth]{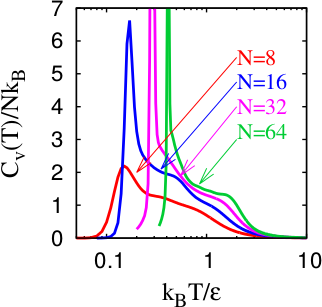}\\
\vspace{7mm}
\includegraphics[width=0.75\columnwidth]{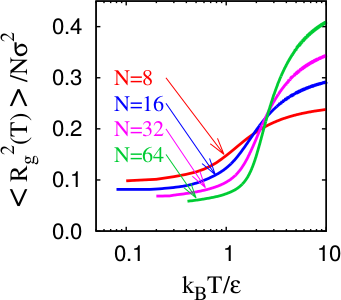}
\caption{ (Top) The specific heat per monomer $C_v(T)/(N k_B)$ as a function of the reduced temperature $k_b T/\epsilon$
for the P model with $N=8,16,32,64$. (Bottom) The reduced mean-square radius of gyration $\langle R^2(T) \rangle / N\sigma^2$ as a function of the reduced temperature $k_b T/\epsilon$
under the same canditions.}
\label{fig:fig1s}
\end{figure}

\begin{figure}
\includegraphics[width=0.72\columnwidth]{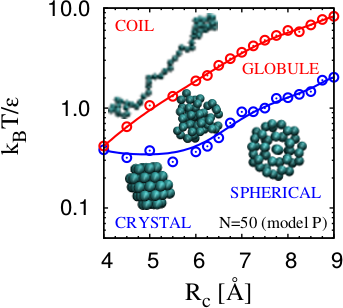}
\caption{Phase diagram of the P model in the reduced temperature $k_BT/\epsilon$ vs interaction range $R_c$ plane
for $N=50$.}
\label{fig:fig2s}
\end{figure}

In Figure \ref{fig:fig1s} we report results obtained from Wang-Landau method applied to model P that are then used to 
draw the phase diagram (see Figure \ref{fig:fig2s}). These include both the specific heat per monomer $C_v(T)/(N k_B)$ 
and the reduced mean-square radius of gyration $\langle R^2(T) \rangle / N\sigma^2$ as a function of the reduced 
temperature $k_b T/\epsilon$.
%%%%%%%%%%%%%%%%%%%%% Fig 1s %%%%%%%%%%%%%%%%%%%%
%\begin{figure}[ht]
%%%%%%%%%%%%%%%%%%%%%%%%%%%%%%%%%%%%%%%%%%%%%%%%
In Figure \ref{fig:fig2s} the phase diagram of the P model in the reduced temperature $k_BT/\epsilon$ vs interaction range $R_c$ plane
is depicted for $N=50$. Upon cooling from a coil configuration, one finds first a transition to a globule, and then
a transition to a crystal. The two transitions merge below   $R_c \approx 6$ \AA to become a direct coil-crystal transition
in agreement with the results by Taylor \textit{et al}. At larger interaction ranges, the crystal structure becomes less
definite more similar to a spherical globule as one could expect on a physical ground.
%%%%%%%%%%%%%%%%%%%%% Fig 2 %%%%%%%%%%%%%%%%%%%%
%\begin{figure}[ht]
%%%%%%%%%%%%%%%%%%%%%%%%%%%%%%%%%%%%%%%%%%%%%%%%
\section{Temperature dependence of the correlation functions for the OPSC model}
\begin{figure}
\includegraphics[width=1.0\columnwidth]{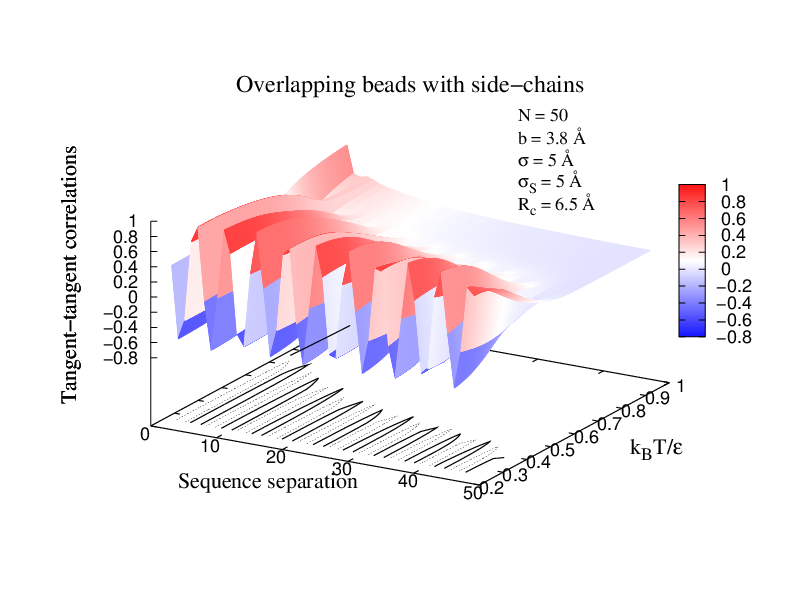}\\
\includegraphics[width=1.0\columnwidth]{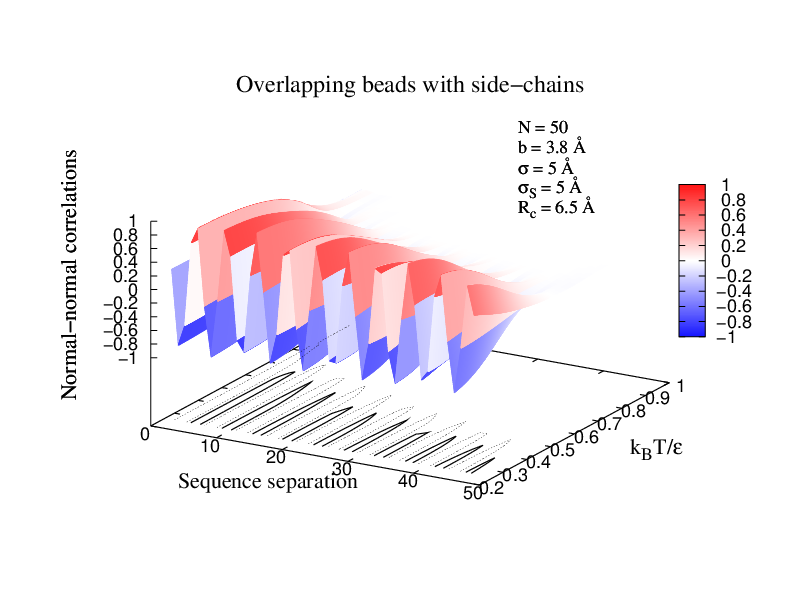}\\
\includegraphics[width=1.0\columnwidth]{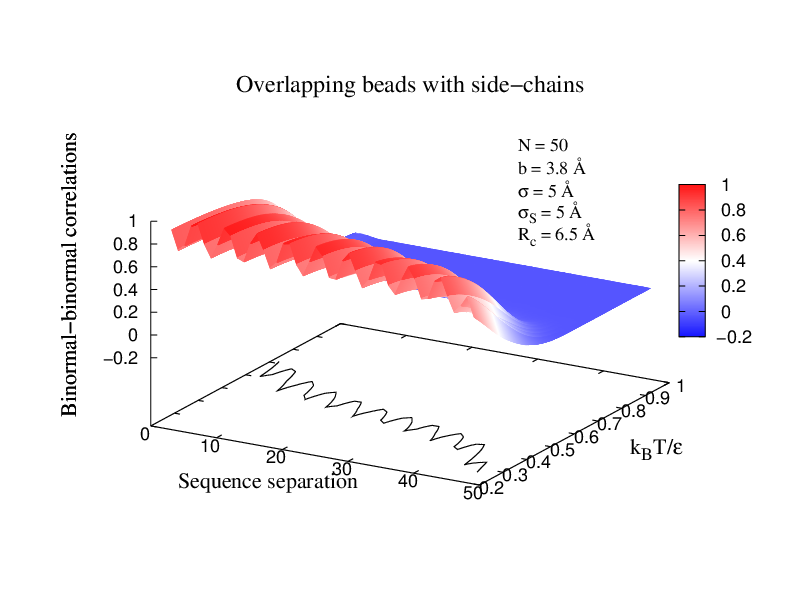}\\
\caption{Three dimensional plot of the tangent-tangent, normal-normal, and binormal-binormal correlation functions as a function of the sequence separation and the reduced temperature $k_B T/\epsilon$. }
\label{fig:fig3s}
\end{figure}

The three correlation functions (tangent-tangent, normal-normal, and binormal-binormal) of the OPSC model are plotted as a function of the sequence separation and of the reduced temperature $k_BT/\epsilon$ in Fig. \ref{fig:fig3s}.  The
transition from exponential to oscillating behavior upon cooling below a critical temperature line is evident in all three cases.
%%%%%%%%%%%%%%%%%%%%% Fig 3  %%%%%%%%%%%%%%%%%%%%
%\begin{figure}[ht]
%%%%%%%%%%%%%%%%%%%%%%%%%%%%%%%%%%%%%%%%%%%%%%%%

\begin{table}
  \caption{Comparison between the ground state energies obtained from the microcanonical Wang-Landau
  and the canonical replica exchange. }
\vspace*{3mm}
\centering
\begin{tabular}{l c c c }
\hline
\hline
$ \lambda=1.05 $ & $N$     & $-E_{\text{gs}}^{\text{WL}}$ & $-E_{\text{gs}}^{\text{RE}}$     \\
&$3$ & $1$                       & $-$                            \\
&$4$ & $3$                       & $-$                            \\
&$5$ & $5$                       & $-$                            \\
&$6$ & $7$                       & $-$                            \\
&$7$ & $10$                       & $-$                            \\
&$8$ & $12$                       & $12$                            \\
&$9$ & $15$                       & $-$                            \\
&$10$ & $18$                       & $-$                            \\
&$12$ & $23$                       & $-$                            \\
&$16$ & $34$                       & $34$                            \\
&$20$ & $46$                       & $-$                            \\
&$32$ & $86$                       & $84$                            \\
&$64$ & $205$                       & $197$                            \\
&$128$ & $451$                       & $-$                            \\
&$256$ & $945$                       & $-$                            \\
\hline
$ \lambda=1.3 $ & $N$     & $-E_{\text{gs}}^{\text{WL}}$ & $-E_{\text{gs}}^{\text{RE}}$     \\
&$3$ & $1$                       & $-$                            \\
&$4$ & $3$                       & $-$                            \\
&$5$ & $5$                       & $-$                            \\
&$6$ & $8$                       & $-$                            \\
&$7$ & $11$                       & $-$                            \\
&$8$ & $13$                       & $13$                            \\
&$9$ & $16$                       & $-$                            \\
&$10$ & $19$                       & $-$                            \\
&$12$ & $27$                       & $-$                            \\
&$16$ & $40$                       & $40$                            \\
&$20$ & $55$                       & $-$                            \\
&$32$ & $103$                       & $102$                            \\
&$64$ & $240$                       & $237$                            \\
&$128$ & $529$                       & $-$                            \\
&$256$ & $1113$                       & $-$                            \\
\hline
$ \lambda=1.5 $ & $N$     & $-E_{\text{gs}}^{\text{WL}}$ & $-E_{\text{gs}}^{\text{RE}}$     \\
&$3$ & $1$                       & $-$                            \\
&$4$ & $3$                       & $-$                            \\
&$5$ & $6$                       & $-$                            \\
&$6$ & $10$                       & $-$                            \\
&$7$ & $13$                       & $-$                            \\
&$8$ & $17$                       & $17$                            \\
&$9$ & $21$                       & $-$                            \\
&$10$ & $25$                       & $-$                            \\
&$12$ & $34$                       & $-$                            \\
&$16$ & $54$                       & $54$                            \\
&$20$ & $73$                       & $-$                            \\
&$32$ & $139$                       & $139$                            \\
&$64$ & $322$                       & $330$                            \\
&$128$ & $690$                       & $-$                            \\
&$256$ & $1450$                       & $-$                            \\
\hline
\hline
\end{tabular}
\end{table}

\begin{acknowledgments}
This work was supported by MIUR PRIN-COFIN2010-2011 (contract 2010LKE4CC). The use of the SCSCF multiprocessor cluster 
at  the Universit\`{a} Ca' Foscari Venezia is greatfully acknowledged. T.X.H. acknowledges support from Vietnam National 
Foundation for Science and Technology Development (NAFOSTED) under Grant No. 103.01-2013.16. R.P. acknowledges support 
from grant P1-0055 of the Slovene Research Agency.
\end{acknowledgments}

\newpage
\bibliography{apsrev}

\end{document}